\documentclass[%
 aip,
 amsmath,amssymb,
 reprint,%
]{revtex4-1}

\usepackage{graphicx}
\usepackage{dcolumn}
\usepackage{bm}

\usepackage[utf8]{inputenc}
\usepackage[T1]{fontenc}
\usepackage{mathptmx}
\usepackage{etoolbox}
\usepackage{amsmath}

\makeatletter
\def\@email#1#2{%
 \endgroup
 \patchcmd{\titleblock@produce}
  {\frontmatter@RRAPformat}
  {\frontmatter@RRAPformat{\produce@RRAP{*#1\href{mailto:#2}{#2}}}\frontmatter@RRAPformat}
  {}{}
}%
\makeatother
\begin{document}

\preprint{AIP/123-QED}

\title[]{Vortex beams with tunable "all-with-visible-light" dye-doped liquid crystal q-plates for broadband application}
\author{Adrián Moya*}
\email{adrian.moya@ua.es}
\author{Adriana R. Sánchez-Montes}%
\affiliation{ 
Instituto Universitario de Física Aplicada a las Ciencias y las Tecnologías, Universidad de Alicante, E-03080 Alicante, Spain.
}%
\author{Sergi Gallego}
\affiliation{ 
Instituto Universitario de Física Aplicada a las Ciencias y las Tecnologías, Universidad de Alicante, E-03080 Alicante, Spain.
}%
\affiliation{%
Departamento de Física, Ingeniería de Sistemas y Teoría de la Señal, Universidad de Alicante, E-03080 Alicante, Spain.
}%
\author{Eva M. Calzado}
\affiliation{ 
Instituto Universitario de Física Aplicada a las Ciencias y las Tecnologías, Universidad de Alicante, E-03080 Alicante, Spain.
}%
\affiliation{%
Departamento de Física, Ingeniería de Sistemas y Teoría de la Señal, Universidad de Alicante, E-03080 Alicante, Spain.
}%

\author{Andrés Márquez}
\affiliation{ 
Instituto Universitario de Física Aplicada a las Ciencias y las Tecnologías, Universidad de Alicante, E-03080 Alicante, Spain.
}%
\affiliation{%
Departamento de Física, Ingeniería de Sistemas y Teoría de la Señal, Universidad de Alicante, E-03080 Alicante, Spain.
}%

\author{Inmaculada Pascual}
\affiliation{ 
Instituto Universitario de Física Aplicada a las Ciencias y las Tecnologías, Universidad de Alicante, E-03080 Alicante, Spain.
}%
\affiliation{%
Departamento de óptica, Farmacología y Anatomía, Universidad de Alicante, E-03080, Alicante, Spain.
}%

\author{Augusto Beléndez}
\affiliation{ 
Instituto Universitario de Física Aplicada a las Ciencias y las Tecnologías, Universidad de Alicante, E-03080 Alicante, Spain.
}%
\affiliation{%
Departamento de Física, Ingeniería de Sistemas y Teoría de la Señal, Universidad de Alicante, E-03080 Alicante, Spain.
}%

\date{\today}

\begin{abstract}
The photoalignment technique for liquid crystal (LC) device fabrication, despite being a well-established method, remains of significant relevance because of its broad applicability. Among its applications, one of particular interest is the generation of structured light, specifically the manufacturing of Pancharatman-Berry (PB) devices, capable of generating optical vortices with angular momentum. In this work, we propose a thorough theoretical and experimental analysis of the optical response of dye-doped liquid crystal (DDLC) devices by examining their performance in terms of tunability and achromaticity across the whole visible spectrum, considering diattenuation effects and how they affect the efficiency of the devices. We experimentally demonstrate the fabrication of photoaligned devices in the visible range with 532 nm laser light and the robust generation of high-quality optical vortices, achieved by a straightforward and accessible technique using a commercial Variable Spiral Plate (VSP), avoiding the need for complex rotational systems or programmable spatial light modulators. Our results demonstrate that diattenuation effects do not prevent the functionality of the devices across the whole visible spectrum and with extended ranges of achromaticity.
\end{abstract}

\maketitle

\section{\label{sec:introduction}INTRODUCTION}

The photoalignment technique  has become a key method for orienting Liquid Crystal (LC) molecules \cite{Yaroshchuk2012, Xi2023, Chigrinov2008} without physical contact, offering benefits over traditional rubbing techniques \cite{Sthr1999, Cui2012}, such as the elimination of permanent substrate surface modifications and potential contamination. Other alternatives, such as nanoparticle alignment \cite{Prakash2022} and surface topography-induced alignment of LCs \cite{Xin2023, Chiou2006} among others \cite{Muravsky2024}, have been proposed over the years. Phototalignment is a non-invasive approach that enables precise local control over LC alignment with the exclusive use of light, facilitating the fabrication of advanced photonic devices. Recent advancements have expanded the applicability of photoalignment in areas like spatial light modulators \cite{Chigrinov2020,Nys2022}, rewritable liquid-crystal-
alignment technology \cite{Muravsky2007}, optical switches \cite{Duan2016}, diffraction gratings \cite{Sun2013}, Fresnel lenses \cite{Huang2012}, etc, demonstrating the versatility of this technique in modern photonic applications. A very appealing application which can take full advantage of photoalignment technologies is the generation of structured light, where a thorough modelling of interplay between retardation and diattenuation might widen the scope of fabrication approaches, as we will show in the paper.

Structured light generation is an area of modern optics with a lot of potential \cite{Forbes2021, Rosales-Guzmn2024, Li2023}, where liquid crystals have 
proven essential in producing beams with complex polarization states, such as radial and azimuthal polarizations. 
These polarization states are characterised by their distinct electric field distributions, which are advantageous in 
applications like high-resolution microscopy \cite{Neil1997, Dong2015}, image edge enhancement \cite{Gao2025}, optical trapping \cite{Yang2021, Kritzinger2022}, optical communications \cite{Trichili2016} and 
optical sensing \cite{Belmonte2015}, among others. The ability to produce beams of this type has been demonstrated using LC-based 
devices such as Pancharatnam-Berry (PB) optical devices \cite{Tian2025, Li2019}, Spiral Phase Plates \cite{Beijersbergen1994}, q-plates \cite{Sit2024, Rubano2019} or Liquid Crystal on 
Silicon spatial light modulators (LCoS) \cite{Grunwald2022, Snchez-Montes2025_1, Snchez-Montes2025}, enabled by the anisotropic properties of LCs to manipulate the 
properties of light. Furthermore, photoalignment techniques make it possible to manufacture devices capable of 
producing structured light output by local and tunable modulation of a polarized beam, enabling the generation of structured light with high 
efficiency \cite{Liu2023, Ko2008}. Depending on the physical principle, large ranges of achromaticity can be obtained in the generation of structured beams, as when illuminating PB devices with broadband light. However, a full exploration of the combination of achromaticity and tunability with diattenuation phenomena still needs to be undertaken.

The photoalignment technique using azo dyes is based on the photoisomerization mechanism of azobenzene derivatives, 
which undergo a reversible cis-trans transformation upon exposure to polarized light. When a layer doped with an azo 
dye is illuminated with linearly polarized light, the dye molecules reorient themselves perpendicularly to the 
polarization direction. The reorientation of adjacent LC molecules is induced by surface anchoring forces \cite{Yaroshchuk2012, Xi2023, Chigrinov2008}. Some compounds, like the Methyl Red (MR) azo dye, are used mixed with LC and they are known as Dye Doped Liquid Crystals (DDLC). Exposure of MR causes its molecules to undergo a brownian effect within the cell in which they 
migrate towards the substrates, where they are adsorbed because of electrostatic interactions at the LC-substrate 
interface. Afterwards, MR undergoes the same cis-trans photoisomerization process as the other azo dyes \cite{Huang2008, Fuh2019}. Most photoalignment azo dyes absorb in the UV, which restricts their use to laboratories with specialized optical setups. By contrast, some azo dyes, like the Methyl Red, enable photoalignment in the visible range, specifically the 532 nm range, but its 
strong absorption leads to diattenuation effects that can compromise the performance of LC devices, thus deserving an in-depth analysis to validate this approach. In a previous paper \cite{Moya2025}, we have already discussed in detail the diattenuation effects on homogeneous liquid crystal cells fabricated via photoalignment. In this paper, we will extend this study into more complex devices, able to generate structured light.

Different approaches can be followed to generate the spatially variant linearly polarized light beams needed to produce complex photoaligned patterns. LCoS devices combined with a quarter-waveplate \cite{Li2019}, scanning systems combined with rotating polarizers \cite{Fuh2019}, combination of adaptive spiral phase plates \cite{Snchez-Lpez2025} or a variable spiral plates\cite{spiralplate} are versatile approaches. If low polarization order light is of interest, then there is no need for complicated setups with mechanically moving elements \cite{Fuh2019} or spatial light modulators \cite{Garca-Martnez2020, Zheng2015}. By spatially varying the polarization of the incident light using a spiral plate \cite{spiralplate}, cylindrical vector beams (CVB) \cite{Zhan2009} can be written onto the substrate with high precision to generate well-known q-plates \cite{Sit2024, Rubano2019}. Devices fabricated in this way can act as geometric phase elements, modulating the light’s orbital angular momentum to generate structured beams. Variations in wavelength-dependent absorption may impact the quality and efficiency of optical vortex generation, yet these effects have not been extensively studied in the literature. In this paper, as we are analyzing exclusively the visible range of the light spectrum, we will use the term “all-with-visible-light” to refer to our manufactured devices.

In this work, we address the various challenges introduced through comprehensive spectral characterisation and robust testing of absorption, achromaticity and tunability of DDLC photoaligned devices, with the objective of fully exploit the potential of visible-spectrum azo dyes in LC photonic devices. The paper is structured as follows. First, in Section \ref{sec:theory}, we show the theoretical development. In Section \ref{sec:sample}, we present the fabrication process and the experimental set-ups. Finally, in Section \ref{sec:results}, we show and discuss the results obtained. 

\section{\label{sec:theory}THEORY}

As discussed in the Introduction, the azo dye employed is Methyl Red, which exhibits absorption within the visible spectrum. Consequently, the manufactured devices display diattenuation effects during operation. 
Therefore, we start developing a theoretical model to analyse a linear retarder with diattenuation using the Jones formalism, and afterwards we will discuss what happens when there is a spatial variation of the neutral lines when generating the well-known vortex beams with orbital angular momentum.
\\
\subsection{Geometric phase and diattenuation}

Throughout this work, we adopt a right-handed coordinates system to describe all vector quantities. In this system, the X and Y axes are aligned with the laboratory’s horizontal and vertical directions, respectively, while the Z axis 
corresponds to the direction of light propagation. Right (left) handedness is defined as clockwise (counterclockwise) sense of direction from the perspective of an observer receiving the light. In general, the device can be modelled as a rotated linear retarder with diattenuation (W) as follows,
\begin{eqnarray}
W(\Gamma, p_{e}, p_{o}, \theta)=R(-\theta)\cdot W_{0}(\Gamma,p_{e}, p_{o})\cdot R(\theta)
\label{eq:rot_ret}.
\end{eqnarray}
, where R is the Jones rotation matrix, $\theta$ is the angle of rotation of the neutral lines, $W_{0}$ is the Jones matrix of a diattenuating retarder with its extraordinary and ordinary lines along the X and Y axes of the reference system, respectively. And $p_{e}$ and $p_{o}$ are the amplitude attenuation coefficients for the extraordinary and ordinary axes.

The R matrix is defined as

\begin{eqnarray}
    R(\theta)=
    \begin{pmatrix}
        cos~\theta & sin~\theta\\
        -sin~\theta & cos~\theta
    \end{pmatrix}
    \label{eq:rotation},
\end{eqnarray}
and the $W_{0}$ matrix can be written as

\begin{eqnarray}
    W_{0}=
    \begin{pmatrix}
        p_{e}e^{-j\varphi_{e}} & 0\\
        0 & p_{o}e^{-j\varphi_{o}}
    \end{pmatrix}
    \label{eq:retard_0},
\end{eqnarray}
 where $e^{-j\varphi_{e}}$ and $e^{-j\varphi_{o}}$ are the phase terms related to the extraordinary and ordinary axes of the retarder. The relation between the phase terms and the retardance $\Gamma$ is given by:

\begin{eqnarray}
    \Gamma =\varphi_{e}-\varphi_{o}=\frac{2\pi}{\lambda}d(n_{e}(\lambda, \alpha)-n_{o}(\lambda))
    \label{eq:diff_phase},
\end{eqnarray}
 where d is the thickness of the device, $n_{e}$ and $n_{o}$ are the extraordinary and ordinary refractive indexes respectively. $n_{e}$ has a dependence with the wavelength $\lambda$ and the rotation angle $\alpha$ of the LC molecules relative to their rest position (along the substrate plane), what is normally called tilt angle. Substituting Eq.~(\ref{eq:diff_phase}) into Eq.~(\ref{eq:retard_0}) results in the following expression:

\begin{eqnarray}
    W_{0}= e^{-j~\varphi_{o}}e^{-j~\Gamma/2}
    \begin{pmatrix}
        p_{e}e^{-j~\Gamma/2} & 0\\
        0 & p_{o}e^{j~\Gamma/2}
    \end{pmatrix}
    \label{eq:retarder_1}.
\end{eqnarray}

Substituting Eq.~(\ref{eq:rotation}) and (\ref{eq:retarder_1}) in Eq.~(\ref{eq:rot_ret}), we obtain:
\begin{widetext}
\begin{equation}
W(\Gamma, p_{e}, p_{o}, \theta)= e^{-j~\varphi_{0}}e^{-j~\Gamma/2}
\begin{pmatrix}
p_{e}e^{-j~\Gamma/2}\cdot cos^{2}\theta+p_{o}e^{j~\Gamma/2}\cdot sin^{2}\theta & \left(p_{e}e^{-j~\Gamma/2}-p_{o}e^{j~\Gamma/2}\right)cos\theta sin\theta\\
\left(p_{e}e^{-j~\Gamma/2}-p_{o}e^{j~\Gamma/2}\right)cos\theta sin\theta & p_{e}e^{-j~\Gamma/2}\cdot sin^{2}\theta+p_{o}e^{j~\Gamma/2}\cdot cos^{2}\theta
\end{pmatrix}
\label{eq:retarder_2}.
\end{equation}
\end{widetext}

Applying some trigonometric identities and considering that the birefringent device is illuminated with circularly polarized light, as usual with PB devices, the electric field at the output can be written as follows:

\begin{widetext}
\begin{equation}
\vec{E}_{out}(\Gamma, p_{e}, p_{o}, \theta)=\frac{e^{-j~\varphi_{o}}e^{-j~\Gamma/2}}{2\sqrt{2}}
\begin{pmatrix}
\left(p_{e}e^{-j~\Gamma/2}+p_{o}e^{j~\Gamma/2}\right)+\left(p_{e}e^{-j~\Gamma/2}-p_{o}e^{j~\Gamma/2}\right)exp(\pm j2\theta)\\
\pm j\left(p_{e}e^{-j~\Gamma/2}+p_{o}e^{j~\Gamma/2}\right)\mp j\left(p_{e}e^{-j~\Gamma/2}-p_{o}e^{j~\Gamma/2}\right)exp(\pm j2\theta)
\end{pmatrix}
\label{eq:retarder_3},
\end{equation}
\end{widetext}
 where in the appearances of '$\pm$' and '$\mp$' the upper (lower) sign is related to incident light polarization right-handed circular, RHC (left-handed circular, LHC).

Very conveniently, Eq.~(\ref{eq:retarder_3}) can be decomposed into the superposition of a RHC and a LHC polarization state. Without any loss of generality and to ease the discussion, we will consider that the SOP for the incident light corresponds to RHC, and then the output light electric field can be expressed as,

\begin{widetext}
\begin{equation}
\vec{E}_{out}(\Gamma, p_{e}, p_{o}, \theta)=\frac{e^{-j~\varphi_{o}}e^{-j~\Gamma/2}}{2\sqrt{2}}\left\{\left(p_{e}e^{-j~\Gamma/2}+p_{o}e^{j~\Gamma/2}\right)
\begin{pmatrix}
    1\\
    j
\end{pmatrix}
+\left(p_{e}e^{-j~\Gamma/2}-p_{o}e^{j~\Gamma/2}\right)exp(j2\theta)
\begin{pmatrix}
    1\\
    - j
\end{pmatrix}
\right\}
\label{eq:retarder_4}.
\end{equation}
\end{widetext}

By rewritting the expression as a function of the average amplitude $\left(p_{avg}=\frac{p_{e}+p_{o}}{2}\right)$ and the average amplitude difference $\left(p_{diff}=\frac{p_{e}-p_{o}}{2}\right)$, we can reformulate Eq.~(\ref{eq:retarder_4}) as:

\begin{widetext}
\begin{eqnarray}
\vec{E}_{out}(\Gamma, p_{e}, p_{o}, \theta)=\frac{e^{-j~\varphi_{o}}e^{-j~\Gamma/2}}{\sqrt{2}}\biggl\{\Bigr[p_{avg}\cdot cos\left(\Gamma/2\right)-j~p_{diff}\cdot sin\left(\Gamma/2\right)\Bigr]
\begin{pmatrix}
    1\\
    j
\end{pmatrix}\nonumber\\
+\Bigr[p_{diff}\cdot cos\left(\Gamma/2\right)-j~p_{avg}\cdot sin\left(\Gamma/2\right)\Bigr]exp(j2\theta)
\begin{pmatrix}
    1\\
    -j
\end{pmatrix}
\biggl\}
\label{eq:retarder_5}.
\end{eqnarray}
\end{widetext}

If there is no diattenuation $\left( p_{avg}=1,  p_{diff}=0\right)$, we obtain the more usual expression:
\begin{widetext}
\begin{eqnarray}
    \vec{E}_{out}(\Gamma, \theta)=\frac{e^{-j\varphi_{o}}e^{-j\Gamma/2}}{\sqrt{2}}\left\{cos\left(\Gamma/2\right)
    \begin{pmatrix}
        1\\
        j
    \end{pmatrix}
    -j~sin\left(\Gamma/2\right)~exp(j2\theta)
    \begin{pmatrix}
        1\\
        -j
    \end{pmatrix}
    \right\}
    \label{eq:reterder_no_diatt}.
\end{eqnarray}
\end{widetext}

Eq.~(\ref{eq:retarder_5}) is central for the goals of the paper, but first it is more instructive to analyse the simplified Eq.~(\ref{eq:reterder_no_diatt}). Looking at Eq.~(\ref{eq:reterder_no_diatt}), the output light from the device is composed of two electromagnetic wavefronts with orthogonal circular polarizations: a first term with RHC polarized light and a second term with LHC light. These two states of polarization, despite being overlapped, are complementary, i.e. they do not interfere with each other. In the second term corresponding to the circularly polarized state complementary to the input state, there is a phase term $exp(j2\theta)$ that does not depend on $\Gamma$, but on the device's relative orientation of its neutral lines. This term is known as the geometric phase, which, unlike the dynamic phase, does not depend on the optical path travelled and the wavelength $\lambda$, thus it has an achromatic character, which is important for the goals of this work.

The retardance $\Gamma$ in Eq.~(\ref{eq:reterder_no_diatt}) can be adjusted, so that one of the two terms cancels, what makes possible to choose between the first, constant, term and the second term with the geometric phase associated. For the case when $\Gamma$ is $180^{\circ}$ (half-wave plate) or an odd multiple: $(2\text{n}+1)\cdot180^{\circ}$ with n an integer, only the geometric phase is produced:

\begin{eqnarray}
    \vec{E}_{out}(\pi, \theta)=\frac{e^{-j\varphi
    _{o}}}{\sqrt{2}}exp(j2\theta)
    \begin{pmatrix}
        1\\
        -j
    \end{pmatrix}
    \label{eq:ret_sin_diat_ret_180}.
\end{eqnarray}

Returning to Eq.~(\ref{eq:retarder_5}), the diattenuation does affect the output beam, but the geometric phase term $\text{exp}(j2\theta)$ itself is unaffected. However, despite the geometric phase's independence, even with $\Gamma=180^{\circ}$, the first term associated to the right-handed circularly polarized light does not disappear: so in contrast with the previous discussion for Eq.~(\ref{eq:reterder_no_diatt}), now the output beam would be a combination of the two terms of different polarization. We cannot cancel any of the overlapped terms, but we can choose a $\Gamma$ that makes minimum the intensity of one of the orthogonal polarizations, as we will demonstrate in Section \ref{sec:results}. The complementary term would contain the geometric phase while the other term would be superimposed light that will reduce the contrast of the image obtained. However, depending on the design for the spatial variations of the angle $\theta$, they do not need to spatially overlap at the plane of interest for the application, as we explain in the last paragraph in the next Subsection.

\subsection{Spatially variant neutral lines devices}

The dependence of the geometric phase on the spatially dependent orientation of the neutral lines allows the creation of different phase elements. In order to analyse this behaviour, we  express the coordinates of the device plane in a polar reference system, where r and $\varphi$ are the radial and azimuthal coordinates, respectively. In the present paper, we are mainly interested in two equivalent situations, which are the ones that can be produced by the VSP:

\paragraph{$\theta(r,\varphi)=\varphi$.} The illuminating wavefront generated through the VSP is azimuthally polarized. The neutral lines orientation generated on the DDLC device through photoalignment points along the radius from the centre of the device. The geometric phase term in the Eqs. in the previous Subsection can be written as $f(\theta=\varphi)=exp(j2\varphi)$, which is equivalent to a beam with orbital angular momentum of topological charge of 2.

\paragraph{$\theta(r, \varphi)=\varphi+\frac{\pi}{2}$.} The VSP generates radially polarized light. The neutral lines generated on the DDLC device points azimuthally and concentrally around the centre of the device, implying that the geometric phase term is written as $f\left(\theta=\varphi+\frac{\pi}{2}\right)=exp( j(2\varphi+\pi))$. Again, this implies that the output beam has a topological charge of 2, but $180^{\circ}$ out of phase compared to case \textit{a}.

\begin{figure}[h!]
\includegraphics[scale = 0.3]{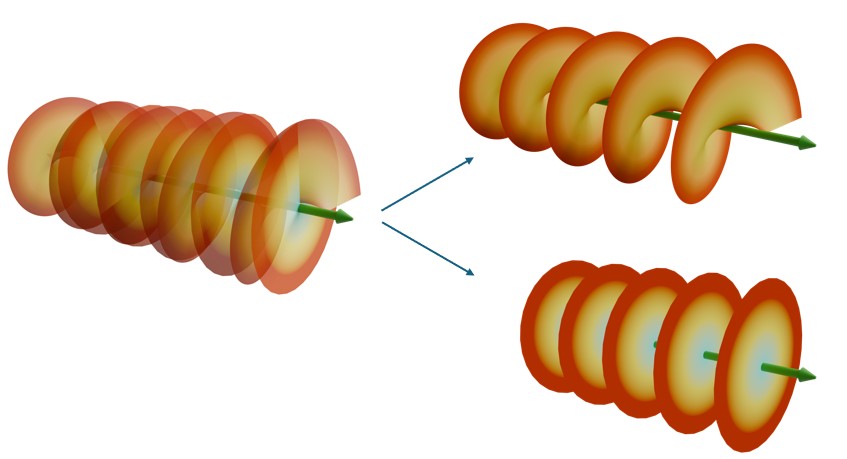}
\caption{\label{fig:figura0} Output beam from the photoaligned device, where two different wavefronts overlap each other: a planar wavefront and a helical wavefront, each of them associated to the two orthogonal circularly polarized states. If the input beam has RHC polarization, then the optical vortex will have LHC polarization, and the planar wavefront RHC polarization.}
\end{figure}

To sum it up, we will produce q-plates able to generate vortex beams with topological charge 2. In Fig.~\ref{fig:figura0}, we show a graphical representation of the overlapped wavefronts when considering Eq.~(\ref{eq:retarder_5}) and (\ref{eq:reterder_no_diatt}): we have the optical vortex associated with the term with the geometric phase, and the planar wavefront associated with the other term of the equation. Since the two wavefronts have complementary circular polarization states, we could filter one of them by a convenient arrangement of a quarter-wave plate and a linear polarizer. In this case, at least 50\% of the energy is lost. In general, what is done is to produce the needed retardance $\Gamma=180^{\circ}$ (or odd multiples) so that the vortex beam is generated with maximum efficiency. Resulting from Eq.~(\ref{eq:retarder_5}), in the presence of diattenuation, even for $\Gamma=180^{\circ}$ (or odd multiples), there is an overlap with the planar wavefront given by the orthogonal polarization term. This spatial overlap, however, disappears in the far-field, usually the regime of observation of vortex beams in applications, where the helical wavefront generates an intensity ring and the planar wavefront corresponds to light in the center, thus not affecting the helical phase along the ring.

\section{\label{sec:sample}SAMPLE PREPARATION AND EXPERIMENTAL SETUP}

Liquid crystal devices incorporating Methyl Red doped LC layers were fabricated via photoalignment technique. The substrates employed consisted of 25 mm × 37.5 mm glass slides coated with a 23 nm layer of Indium Tin Oxide (ITO). As a first step, the substrates underwent a surface treatment, which included a cleaning process with acetone (Merck, Darmstadt, Germany) and an ultrasonic bath for a duration of 15 minutes (Equipos Clínicos, Jaen, Spain), followed by drying in an oven (Fisher Scientific, Madrid, Spain) at $120^{\circ}C$ for 20 minutes. Subsequently, the surface was exposed to UV ozone treatment for 15 minutes using an Ossila UV Ozone Cleaner (Ossila, Sheffield, UK). The cells were then assembled using two substrates and silica spacers (5.5 \textmu m, Whitehouse Scientific, Chester, UK) to define the thickness of the devices. The selected thickness value ensured that retardances exceeding $2\pi$ could be obtained. The spacers were deposited via spin-coating, and two sides of the cell were sealed with a UV adhesive (Norland Optical Adhesive 61, Edmund Optics, Barrington, USA) \cite{Moya2025}.

A liquid crystal mixture containing 99 wt\% E7 (Nematel, Mainz, Germany) and 1 wt\% Methyl Red MR (Merck, Darmstadt, Germany) was introduced into the cells through capillary action. To optimize filling and ensure uniform distribution of the LC, the process was conducted at approximately $65^{\circ}C$, which is the temperature of the isotropic phase of the LC, where the molecules lose its birefringence, and it behaves like a liquid. This isotropic condition allows for the simultaneous photoalignment of both cell surfaces, using a method known as double-sided photoalignment \cite{Lin2007}. The thickness of our device was measured using a known interferometric method that we recently used in our last work \cite{Moya2025}, obtaining a value of 7.4 \textmu m.

\begin{figure}[h!]
\includegraphics[scale = 0.13]{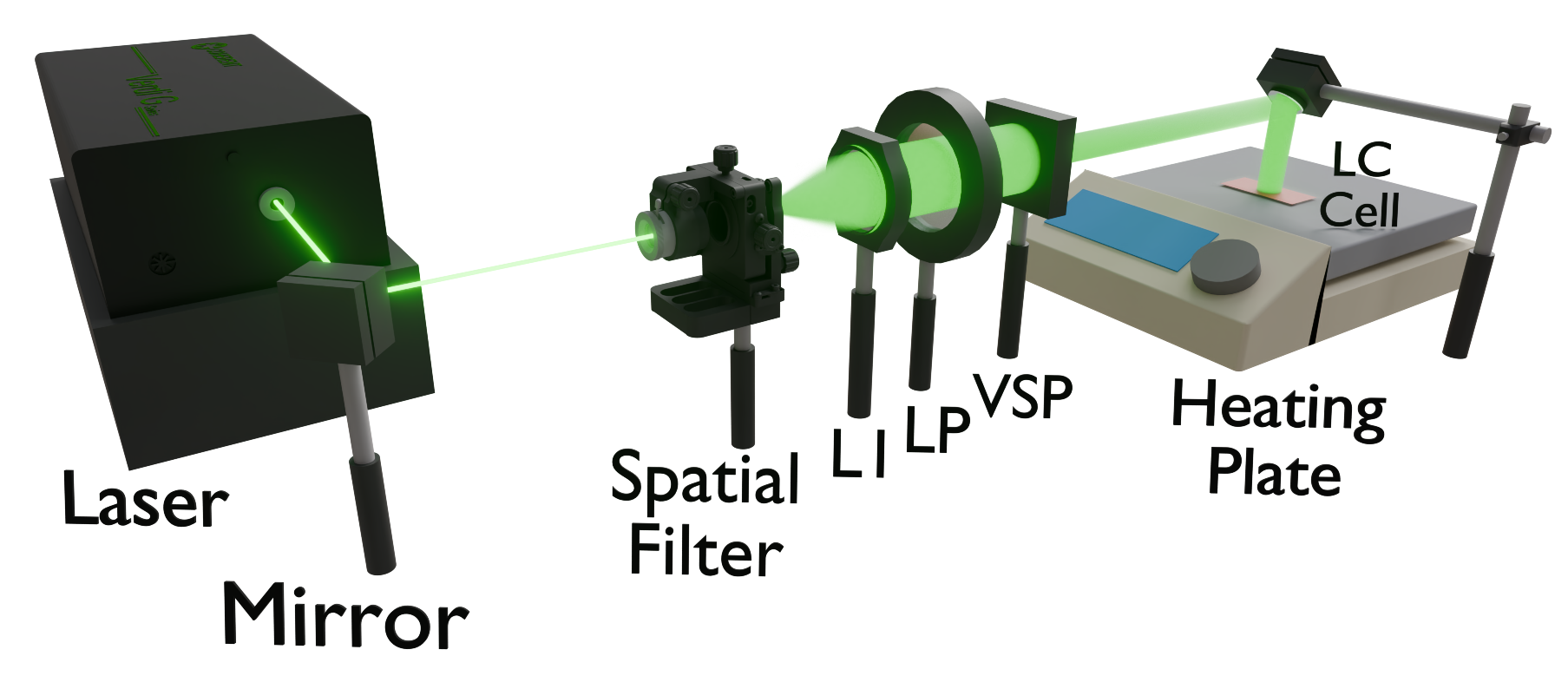}
\caption{\label{fig:figura1} Photoalignment setup.}
\end{figure}

The photoalignment set-up is shown in Fig. \ref{fig:figura1}. First, a 532 nm laser beam (Verdi-V5, Coherent, Copenhagen, Denmark) is expanded and collimated by a combination of spatial filter and a lens (L1). Then, we obtain linearly polarized light along the vertical axis of the laboratory through a linear polarizer (LP), which passes through the Variable Spiral Plate (VSP, ArcOptix, $\mathrm{Ne\hat{u}chatel}$, Switzerland) \cite{spiralplate}, capable of generating structured light: specifically radial and azimuthally polarized light. Finally, this structured polarized light impinges perpendicular to the LC cell, which is at a temperature of 65 °C in order to keep it in an isotropic state to ensure a correct photoalignment in both areas of the substrates. The photoaligned area of our devices is 1 cm in diameter, since that is the size of the aperture of the VSP. With this assembly, we can generate photoaligned devices with an azimuthal (and/or radial) distribution of the LC molecules.

\begin{figure}[h!]
\includegraphics[scale = 0.14]{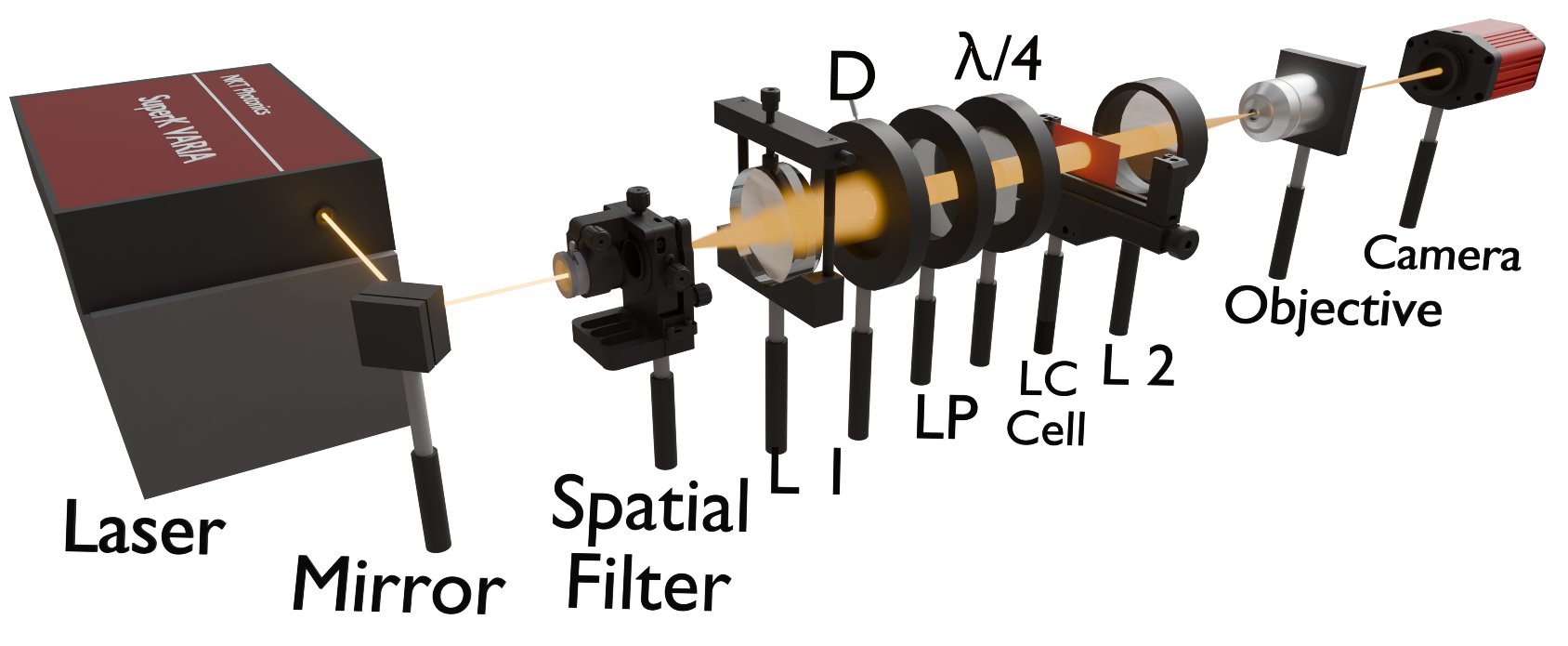}
\caption{\label{fig:figura2} Characterization setup.}
\end{figure}

Fig. \ref{fig:figura2} shows the set-up for the characterisation of the manufactured devices. In this case, the cell was illuminated by a supercontinuum laser (NKT Compact, NKT Photonics, Birkerod, Denmark) with a wavelength selector (SuperK Varia, NKT Photonics, Birkerod, Denmark). The laser was operating at its maximum spectral bandwidth (450 – 850 nm) covering all the visible spectrum. First, the beam is collimated with a combination of a spatial filter and a lens L1 (focal = 70 mm) and its diameter is adjusted with the diaphragm (D) to determine the area to be analysed (around $0.8~\text{cm}^{2}$). As established in Section \ref{sec:theory}, to produce the vortex beam, we have to illuminate the device with circularly polarized light. For that we use a combination of a linear polarizer (LP) with its transmission axis positioned at $0^{\circ}$ and a quarter-wave plate retarder with its neutral lines oriented at $45^{\circ}$ to the laboratory’s horizontal. Finally, at the focal plane of lens L2 (focal = 200 mm), we are able to obtain the far field pattern of the beam. To capture the optical vortices at the lens focal plane, a microscope objective is used, allowing an image of adequate size to be captured by the camera sensor (CS126CU CMOS colour Camera, Thorlabs, New Jersey, USA).

\begin{figure}[h!]
\includegraphics[scale = 0.42]{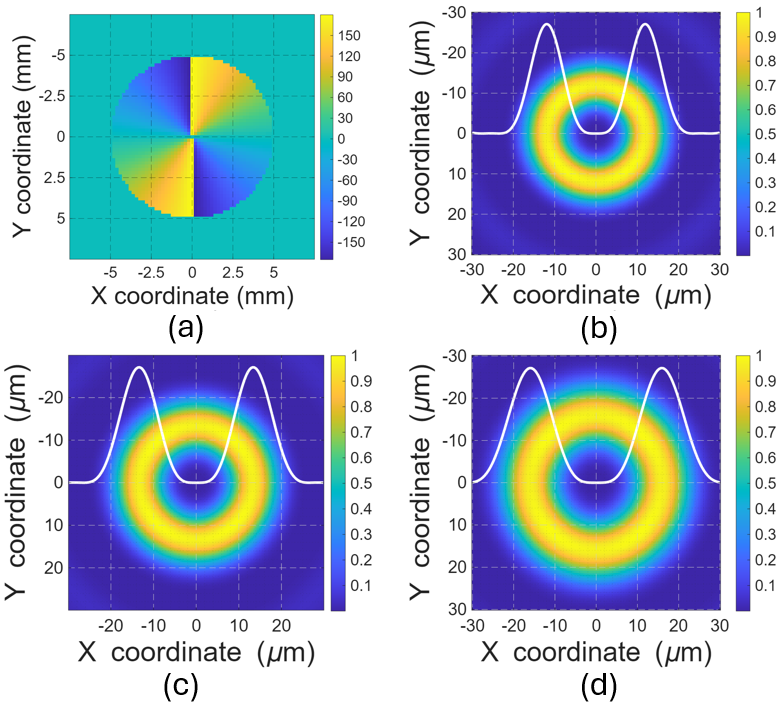}
\caption{\label{fig:figura2.5} Laboratory scaled simulation of the optical vortices generated by our photoaligned q-plate, where (a) shows the phase profile generated by a q-plate with topological charge 2 and (b-d) are the optical vortices obtained when the incident wavelegnth is 473 nm, 532 nm and 633 nm, respectively, at the focal plane of lens L2.}
\end{figure}

Once the setup has been established, before showing the experimental results, we can theoretically simulate the optical vortices that we should obtain in the focal plane of L2. Knowing that the focal length is 200 mm and the illuminated area of the photoaligned q-plate is 0.8 $\text{cm}^{2}$ approximately, we can simulate the behaviour of the device for the three wavelengths we will study along the paper: 473 nm, 532 nm and 633 nm. In Fig.~\ref{fig:figura2.5}, in subfigure (a), we can see the phase profile of the helical wavefront at the exit of the DDLC cell, which corresponds to a geometric phase with a topological charge of 2 when RHC light is incident. Subfigures (b-d), show the theoretical optical vortices that we should obtain at the focal plane of L2 with lengths expressed in laboratory coordinates for the sake of comparison with the experimental images in the next Section. As we can observe, its diameter increases slightly with the wavelength, and the ring diameter is about 20 \textmu m diameter.

It is important to note that our experimental setup introduces a certain degree of tolerance, particularly in the positioning of the magnifying objective at the focal plane of L2. As a result, slight discrepancies between the size of the optical vortices are to be expected between the experimental results and the simulated ones.

\section{\label{sec:results}RESULTS}

Since the results obtained from the radially and azimuthally photoaligned devices are similar, in this Section we will only analyse the azimuthal patterns, as we found that it is the most efficient configuration with the current materials and VSP setup available in our laboratory. The study of the device was carried out in terms of voltage response and wavelength-dependent behaviour, with a focus on the voltage values that make this behaviour independent of the wavelength across the visible spectrum (achromaticity).

\begin{figure}[h!]
\includegraphics[scale = 0.65]{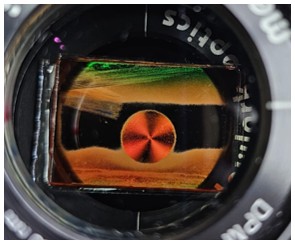}
\caption{\label{fig:figura3} q-plate of 7.4~\textmu m thickness between crossed polarizers.}
\end{figure}

In Fig. {\ref{fig:figura3}}, we show the pattern of a device azimuthally photoaligned, when placed between two crossed polarizers and ambient illumination. It can be appreciated with great definition the circular aperture of the photoalignment process, where we can clearly see the four orthogonal nulls of intensity, what implies a good photoalignment quality.

Let us first provide an estimation of the magnitude of the planar wavefront which is added to the vortex beam when there is diattenuation. For that, we have to check the optimal value of $\Gamma$ so that we obtain maximum intensity $\text{I}_{max}$ for the geometric phase term and minimum intensity $\text{I}_{min}$ for the first term. From Eq.~(\ref{eq:retarder_5}), the intensity associated with each term can be calculated as:

\begin{eqnarray}
    I_{1}=\left|\left(\frac{p_{e}+p_{o}}{2}\right)cos(\Gamma/2)-j\left(\frac{p_{e}-p_{o}}{2}\right)sin(\Gamma/2)\right|^{2}
    \label{eq:energia_term_1}
\end{eqnarray}

\begin{eqnarray}
    I_{2}=\left|\left(\frac{p_{e}-p_{o}}{2}\right)cos(\Gamma/2)-j\left(\frac{p_{e}+p_{o}}{2}\right)sin(\Gamma/2)\right|^{2}
    \label{eq:energia_term_2},
\end{eqnarray}
 where $I_{1}$ and $I_{2}$ are the intensities associated with the first and second terms of Eq~(\ref{eq:retarder_5}), respectively.

After some manipulations and simplifying these expressions with the use of trigonometric identities, we obtain:

\begin{eqnarray}
    I_{1}=\frac{1}{4}\left(p_{e}^{2}+p_{o}^{2}+2p_{e}p_{o}cos\Gamma\right)
    \label{eq:ener_simpl_term_1}
\end{eqnarray}

\begin{eqnarray}
    I_{2}=\frac{1}{4}\left(p_{e}^{2}+p_{o}^{2}-2p_{e}p_{o}cos\Gamma\right)
    \label{eq:ener_simpl_term_2}.
\end{eqnarray}

It is easy to notice that the term that minimizes Eq.~(\ref{eq:ener_simpl_term_1}) and maximizes Eq.~(\ref{eq:ener_simpl_term_2}) is $\Gamma=(2n+1)\cdot 180^{\circ}$, with n an integer.

\begin{eqnarray}
    I_{1,min}=\frac{1}{4}\left(p_{e}-p_{o}\right)^{2}
    \label{eq:ener_min_term_1}
\end{eqnarray}

\begin{eqnarray}
    I_{2,max}=\frac{1}{4}\left(p_{e}+p_{o}\right)^{2}
    \label{eq:ener_max_term_2}.
\end{eqnarray}

So, if we consider Eq.~(\ref{eq:retarder_5}) for odd multiples of $180^{\circ}$ and for a q-plate with azimuthal orientation of the neutral lines illuminated with RHC polarized light, the output wavefront is:

\begin{eqnarray}
    \vec{E}(180^{\circ}, \varphi)=\frac{e^{-j~\varphi_{o}}}{2\sqrt{2}}\left[p_{diff}
    \begin{pmatrix}
        1\\
        j
    \end{pmatrix}
    +  p_{avg}~e^{j~2\varphi}
    \begin{pmatrix}
        1\\
        -j
    \end{pmatrix}
    \right]
    \label{eq:calc_energia_1}.
\end{eqnarray}

We see that the quotient of both intensities in Eq.~(\ref{eq:ener_min_term_1}) and (\ref{eq:ener_max_term_2}) gives the relative weight of the planar wavefront:

\begin{eqnarray}
    R_{opt}=\left(\frac{p_{e}-p_{o}}{p_{e}+p_{o}}\right)^{2}
    \label{eq:energy_weight}.
\end{eqnarray}

Using the amplitude absorption coefficients we measured in our last work\cite{Moya2025} for a device of 8.4 \textmu m thickness (see Table 2 in Moya et al.\cite{Moya2025}) and applying Eq.~(\ref{eq:energy_weight}), we obtain the corresponding values in Table \ref{tab:table0} for $R_{opt}$. Note that these values are comparable with the 7.4~\textmu m device because the thickness difference is small, so the absorption values will be similar.

Therefore, with Table \ref{tab:table0}, we show that diattenuation effects might have a very small influence on the quality of the vortex beams generated with the "all-with-visible-light" approach. In particular, we confirm that at longer wavelengths, red part of the spectrum, where no diattenuation exists, the value for $R_{opt}$ is almost zero. Diattenuation increases as wavelength decreases. However, even in the blue part of the spectrum, at 473 nm, we see that $\text{R}_{\text{opt}}$ is still very small, close to 1\%. In the next subsections we proceed with a more detailed analysis of the tunability and achromaticity achievable with the DDLC fabricated device.

\begin{table}[h!]
\caption{\label{tab:table0} Absorption coefficients from Moya et al.\cite{Moya2025} and the intensity ratio $R_{opt}$ between planar wavefront and vortex beam, where $p_{e}$ and $p_{o}$ correspond to $p_{x}$ and $p_{y}$ used in that paper, respectively.}
\begin{ruledtabular}
\begin{tabular}{cccc}
$\lambda$(nm)&$p_{e}$ &$p_{o}$&$R_{opt}$ (\%)\\
\hline
473 & 0.62 & 0.79 & 1.45\\
532 & 0.70 & 0.84 & 0.83\\
633 & 0.87 & 0.91 & 0.05\\
\end{tabular}
\end{ruledtabular}
\end{table}

\subsection{\label{tunability}Tunability}

As discussed in Section \ref{sec:theory}, the state where the geometric phase of the device generates more efficiently an optical vortex is reached when $\Gamma$ is an odd multiple of $180^{\circ}$. Experimentally, to obtain an optical vortex using the device, it is necessary to apply a certain voltage to the sample, in order to reach this $\Gamma$ value. A voltage sweep is performed to determine all the $180^{\circ}$ phase shift multiples produced by the device. During this process, the appearance of a well-defined optical vortex ring indicates that a retardance close to $180^{\circ}$ has been achieved. In practice, there is a tolerance range and it is not necessary to have a $\Gamma$ value of exactly $180^{o}$: the optical vortex will appear with a good contrast within a voltage interval. 

Furthermore, to evaluate the tunability of the device across the visible spectrum, three primary wavelengths were chosen to have a proper sampling: red (633 nm), green (532 nm) and blue (473 nm). They correspond as well to laser wavelengths typically found in many laboratories. 

 For the results, we have selected two optical vortices with the best contrast, because in the case of light of 532 nm and 473 nm, our device was able to generate a total of three and four optical vortices, but with very short voltage intervals. We note that the two more robust vortices appeared at the larger voltages. In the case of the 633 nm wavelength, two optical vortices were achievable. In Table $\text{\ref{tab:table1}}$, we show these results for the two voltage intervals studied. We note that voltages are below 5 V in any case. The threshold voltage, when the molecules of LC E7 in our devices start to be affected by the electric field, is around 1 V. 

\begin{figure}[h!]
\includegraphics[scale = 0.65]{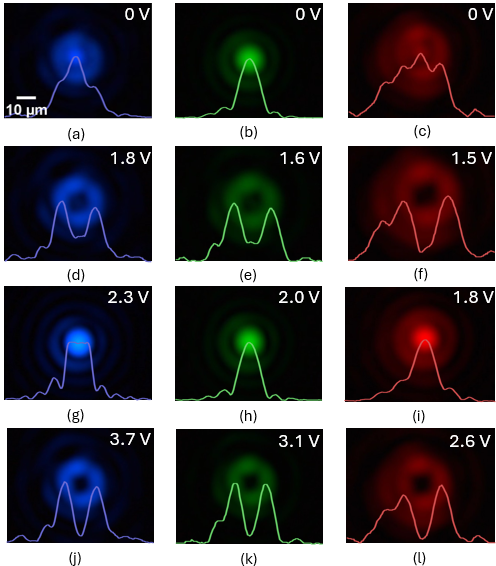}
\caption{\label{fig:figura4} Intensity distributions together with horizontal cut across the centre of the cell for 473 nm, 532 nm and 633 nm, respectively at the first, second and third columns. In each subfigure, we show the applied voltages related to off-state (first row), second-to-last vortex (second row), intermediate planar wavefront (third row) and last vortex (fourth row).}
\end{figure}

 In Fig. \ref{fig:figura4}, we present the intensity distributions of our device under the three different wavelengths. With a horizontal cut across the centre, we can appreciate the profile in the light distribution. The actual size of the patterns of the focal plane of lens L2 can be deduced from the 10~\textmu m scale bar included in the image (a) and they are consistent with the sizes provided in the simulated images in Fig.~\ref{fig:figura2.5}.

 \begin{table}[h!]
\caption{\label{tab:table1}Tunability: Voltage intervals where the vortex beam shows the best contrast for each of the three RGB wavelengths. $\triangle{V}_{1}$ and $\triangle{V}_{2}$ correspond to the ocurrences at the two larger voltages (described in the text).}
\begin{ruledtabular}
\begin{tabular}{ccc}
$\lambda$(nm)&$\bigtriangleup V_{1}$(V)&$\bigtriangleup V_{2}$(V)\\
\hline
473 & 1.7-1.8 & 3.4-4.1\\
532 & 1.5-1.7 & 2.9-3.6\\
633 & 1.3-1.5 & 2.4-3.1\\
\end{tabular}
\end{ruledtabular}
\end{table}

As shown in Fig. \ref{fig:figura4}, each row represents a different state of the total wavefront generated by our device.  As we have analysed in Eq.~(\ref{eq:retarder_5}), depending on the retardance $\Gamma$ and, therefore, the applied voltage, we will have different combinations where the planar and helical wavefronts will have differents weights. In the first row (Fig.~\ref{fig:figura4} (a-c)), we apply a voltage of 0 V, which means that we are not manipulating the weight of each term, and depending on the thickness and wavelength, we will have different situations. For example, at 532 nm, there is no optical vortex, and the dominant term of Eq.~(\ref{eq:retarder_5}) is the first, corresponding to the planar wavefront. On the other hand, for 473 nm and 633 nm, we can see that there is a combination of both terms, so both wavefronts are overlapped. Looking at the intensity profile, we can deduce that in the case of 633 nm, the two terms have similar weights, while for 473 nm, the planar wavefront term is larger. Next, we modified the voltage, trying to achieve situations where one of the terms dominates over the other. In the second row (Fig.~\ref{fig:figura4} (d-f)), we show the first of the optical vortices described in Table \ref{tab:table1}. In this case, the optical vortex can be clearly observed. It is interesting to notice the better contrast obtained for 633 nm with respect to the 532 nm and 473 nm cases. This is probaby due to the absence of diattenuation at 633 nm, as shown in Table \ref{tab:table0}. Therefore, the residual planar wavefront in Eq.~(\ref{eq:calc_energia_1}), given by the weight of $p_{diff}$, is totally cancelled. In the third row (Fig.~\ref{fig:figura4} (g-i)), we have a state where the planar wavefront dominates. And finally, in the fourth row (Fig.~\ref{fig:figura4} (j-l)), we present the second optical vortex, corresponding to the wider voltage ranges presented in Table~\ref{tab:table1}. It is interesting to note that in this case, the contrast is very good for all three wavelengths. These results agree with the theoretical results described in Table \ref{tab:table0}, since the effects of diattenuation affect at most 1.45\% in the case of blue light (473 nm). In addition, the sizes of the optical vortices and the increase in diameter with larger wavelengths are consistent with those obtained in the simulation in Fig.~\ref{fig:figura2.5}, which validates our experimental methodology to manufacture this type of devices.

From a theoretical point of view, we can justify that the voltages required to reach the $180^{\circ}$ phase shift odd multiples decrease as the wavelength increases (as happens at each of the rows in Fig.~\ref{fig:figura4}), analysing the relationship between the physical variables in the retardance equation, described by Eq.~(\ref{eq:diff_phase}). The specific values of the extraordinary and ordinary refractive indices for E7 (the LC used in this work) can be obtained from Moya et al\cite{Moya2025}, and more generally from Li et al\cite{Li2005}. Looking at Eq.~(\ref{eq:diff_phase}), $\Gamma$ is inversely proportional to the wavelength $\lambda$ and proportional to the thickness d and the difference between the refractive indices ($\triangle{n}$), the latter depending on the voltage. In parallel aligned devices, the maximum value for $\triangle{n}$, thus also for $\Gamma$, occurs at 0 Volts and then decreases with the increase in the voltage. Therefore, for a device of thickness d and considering a retardance value $\Gamma$, we need to decrease the voltage as the wavelength increases. This justifies the results shown in Table \ref{tab:table1}, where the voltage values for the device are smaller when characterised using light of 633 nm. 

\begin{figure}[h!]
\includegraphics[scale = 0.37]{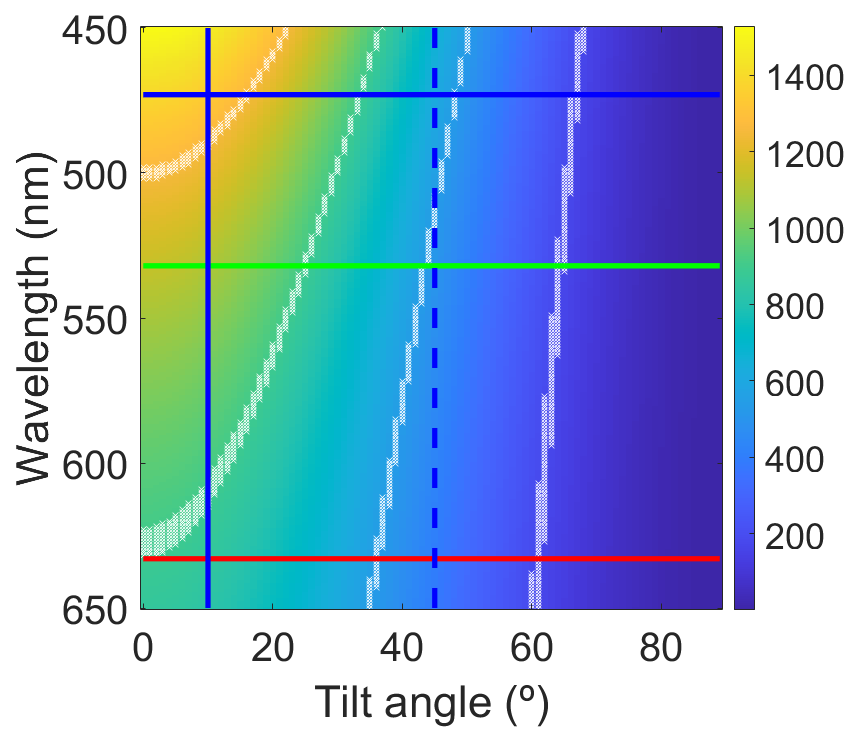}
\caption{\label{fig:figura5.5} Colour plot of the retardance vs tilt angle for the visible spectrum for the 7.4 \textmu m cell fabricated in the paper. The horizontal lines marks the three primary wavelengths studied (473 nm, 532 nm and 633 nm), while the vertical lines marks the two tilt angles ($10^{\circ}$ and $45^{\circ}$) studied in the Achromaticity Subsection.}
\end{figure}

With Eq.~(\ref{eq:diff_phase}), we can also predict the number of optical vortices we can obtain depending on the wavelength $\lambda$ used. In the 2-dimensional colour plot in Fig.~\ref{fig:figura5.5}, the retardance values $\Gamma$ are shown as a function of the visible wavelengths (Y-axis) and the tilt angles (X-axis). Note that in parallel aligned LC devices, as our DDLC cell, the tilt angle increases (non-linearly) with the increase in the applied voltage\cite{Moya2025}. We can also appreciate the white lines that correspond to the values of $\Gamma$ equal to $(2\text{n}+1)\cdot 180^{\circ}$. Horizontal lines are overlapped for the 3 wavelengths analysed in this work (473/532/633 nm). Also, vertical lines at tilt voltages $10^{\circ}$ and $45^{\circ}$ are indicated, which will be used in the next Subsection. Following these theoretical results, for our device of 7.4~\textmu m, we should obtain two optical vortices for 633 nm, three for 532 nm and four for 473 nm, which agrees with the experimental results, discussed when introducing Table~\ref{tab:table1} and shown in Fig.~\ref{fig:figura5.5}. In addition, with this figure, we can appreciate the approximate theoretical retardance ranges of the device: $5\pi$ for 633 nm, $7\pi$ for 532 nm and $8\pi$ for 473 nm. By analogy with our experimental results and the voltage ranges used, we can deduce that the last optical vortex shown in Fig.~\ref{fig:figura4} (fourth row) for each wavelength corresponds to a retardance $\Gamma$ of $180^{\circ}$ (first white line from the right) and the other one appearing at a lower voltage (second row in Fig.~\ref{fig:figura4}) to $\Gamma$ equal to $540^{\circ}$ approximately (second white line from the right in Fig.~\ref{fig:figura5.5}). In Fig.~\ref{fig:figura5.5}, we can also appreciate more clearly the effect we have just discussed about the dependence of $\Gamma$ with the tilt angle (voltage) and the wavelength: so for a fix value of retardance, the tilt angle decreases when the wavelength increases.

\begin{figure}[h!]
\includegraphics[scale = 0.8]{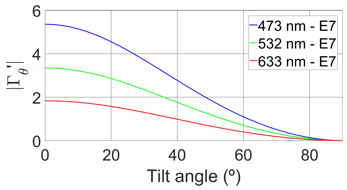}
\caption{\label{fig:figura6} Tilt angle dispersion of the retardance $\Gamma$ as a function of the tilt angle of LC molecules for 7.4~\textmu m.}
\end{figure}

As we have already explained, $\Gamma$ has a dependence on the wavelength $\lambda$ and the tilt angle $\alpha$ (angle between the long axis of the LC molecule and the plane of the substrate), so we can study the evolution of the retardance as a function of $\lambda$ or $\alpha$. In Fig. \ref{fig:figura6} we can see the relationship between the tilt angle related dispersion $\left(\mid\Gamma'_{\alpha}\mid=\mid\frac{\partial\Gamma}{\partial\alpha}\mid\right)$ and the tilt angle. This figure describes how fast the retardance changes for three different wavelengths along the tilt angles, i.e. as a function of the applied voltage. As the tilt angle increases, the rate of change in the dispersion becomes progressively slower, which explains why the variation in retardance decreases at higher voltages and why the second singularity can be observed over a larger range of voltages, as we obtained in the results from Table~\ref{tab:table1} and Fig.~\ref{fig:figura4}. So, from these results and from the tunability point of view, we can assume that the optical vortices obtained with a retardance of $180^{\circ}$ are preferable, because they are more stable (wider voltage range) and show a better contrast. Note that in Fig.~\ref{fig:figura6}, the profile for each wavelength is basically the same apart from a constant scale factor; with values of 0.36 and 0.64 for 633 nm and 532 nm when calculated with respect to the dispersion at 473 nm.

\subsection{\label{achromaticity}Achromaticity}

For the study of the achromaticity of the device, various sweeps will be performed for the three wavelengths considered. Before presenting the results, it is important to define the characterization criteria employed in this study. The images were captured using a 12-bit depth camera, allowing measurements across 4096 greyscale levels, where 0 corresponds to absolute black and 4096 to pure white. Based on this, any greyscale value exceeding 1024 in the singularity area will be considered as a deviation from black, and thus, the optical vortex will no longer be regarded as exhibiting a true singularity. Obviously, a lower greyscale value could also be chosen. However, with the threshold of 1024 we observe that the centre is still very dark, thus it is a reasonable criteria.

\begin{figure}[h!]
\includegraphics[scale = 0.65]{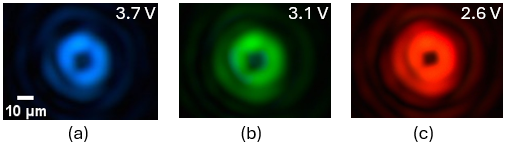}
\caption{\label{fig:figura7} Optical vortices centered at 473, 532 and 633 nm wavelength, with the illumination spectral bandwidth of (a) 65 nm, (b) 90 nm and (c) 100 nm (as shown in Table \ref{tab:table2}).}
\end{figure}

In the first set of measurements, for each of the three wavelengths (473, 532 and 633 nm), the voltage is fixed at the value for which the optical vortex exhibits the maximum contrast. These voltages correspond to the middle value shown in Table \ref{tab:table1} for $\triangle{V}_{2}$ at each of the wavelengths. Subsequently, the spectral bandwidth is gradually increased in order to evaluate the spectral response of the device. Table \ref{tab:table2}  presents the assigned voltage together with the wavelength ranges over which the singularity continues to meet the previously defined criterion. Additionally, Fig. \ref{fig:figura7} shows the optical vortices related to the maximum spectral bandwidths $\triangle{\lambda}$ indicated in Table \ref{tab:table2}, and for each of the three wavelengths. We can appreciate that the three vortices show a good contrast with a dark center, specially for 633 nm.

\begin{table}[h!]
\caption{\label{tab:table2}Achromaticity: Spectral bandwidths within which the optical vortex singularity shows a good contrast at each of the three wavelengths.}
\begin{ruledtabular}
\begin{tabular}{ccc}
$\lambda$(nm)&Voltage (V)&$\bigtriangleup \lambda$ (nm)\\
\hline
473 & 3.7 & 65\\
532 & 3.1 & 90\\
633 & 2.6 & 100\\
\end{tabular}
\end{ruledtabular}
\end{table}

\begin{figure}[h!]
\includegraphics[scale = 0.8]{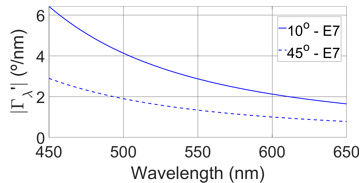}
\caption{\label{fig:figura8} Chromatic dispersion of the retardance $\Gamma$ as a function of the tilt angle of LC molecules.}
\end{figure}

Our results are supported by theoretical calculations, where in Fig.~$\ref{fig:figura8}$, the relationship between the chromatic dispersion of the retardance $\left(\mid\Gamma'_{\lambda}\mid=\mid\frac{\partial\Gamma}{\partial\lambda}\mid\right)$ with the wavelengths is shown for two tilt angles. In this figure, we can clearly see how the chromatic dispersion is lower for higher values of $\lambda$ and higher tilt angles (voltages), justifying that the best achromatic behaviour corresponds to the situation where the device is applied larger voltages and illuminated by light at 633 nm, as demonstrated experimentally in Table \ref{tab:table2}. It is also interesting to notice that the two curves for different tilt angles have almost the same profile apart from a scale factor of 2.15.

We see that the wide wavelength ranges validate the usage of the q-plates not only with laser light but also with broadband illumination from, for example, LED light sources.
\newpage
\section{\label{sec:conclusions}CONCLUSIONS}

This work demonstrates the successful fabrication of devices photoaligned with visible light (532 nm) based on MR doped LC, capable of generating optical vortices through a simple photoalignment setup, which employs a commercially available Variable Spiral Plate to induce radial or azimuthal alignment patterns. 

The optical performance of the devices was thoroughly characterized across the entire visible spectrum, including wavelengths where the azo dye MR shows strong absorption and diattenuation. Remarkably, the device was able to produce structured light even in these, a priori, non-favourable spectral areas, indicating that such absorption does not prevent practical functionality. Novel analytical expressions were developed highlighting the role of diattenuation in geometric phase-based devices, and enabling to quantify its effect through the $\text{R}_{\text{opt}}$ magnitude.

Furthermore, the device's characterization was evaluated in terms of tunability and achromaticity, demonstrating reliable and consistent performance, which validated the proposed fabrication method's versatility and reliability. Both, experimentally and theoretically, we have shown that, in case of various achievable $180^{\circ}$ retardance odd multiples, better results in terms of voltage tunability robustness and achromaticity are produced for the one produced at a larger voltage. The effects of diattenuation do not greatly affect the performance of these "all-with-visible-light" devices, where the worst case scenario is with 473 nm, where the contrast is reduced by 1.45\% ($\text{R
}_{\text{opt}}$ magnitude), which is a small degradation of the contrast. These findings provide valuable insight into the design of ultrathin geometric phase optical elements for the generation of structured light using photoalignment.

\section{ACKNOWLEDGEMENTS}
This work was supported by the Grant CIPROM/2024/90 funded by Generalitat Valenciana (Spain), Grant PID2023-148881OB-I00 funded by MICIU/AEI/ 10.13039/501100011033 and by “ERDF/EU” (Spain) and Grant PID2024-161610OB-I00 funded by MICIU/AEI/ 10.13039/501100011033 and by “ERDF/EU”. ARS-M is grateful to the “Generalitat Valenciana” for the grant (GRISOLIAP/2021/106).

\section{AUTHOR DECLARATIONS}
\subsection{Conflict of Interest}

The authors have no conflicts to disclose.

\subsection{Author Contributions}

\textbf{Adrián Moya}: Conceptualization (equal); Data curation (equal); Formal analysis (equal); Investigation (equal); Methodology (equal); Software (equal); Validation (equal); Visualization (equal); Writing-original draft (equal). \textbf{Adriana R. Sánchez-Montes}: Methodology (equal); Software (equal). \textbf{Sergi Gallego}: Funding acquisition (equal); Resources (equal); Writing-review \& editing. \textbf{Eva M. Calzado}: Conceptualization (equal); Methodology (equal); Project administration (equal); Resources (equal); Supervision (equal); Writing-review \& editing (equal). \textbf{Andrés Márquez}: Conceptualization (equal); Formal Analysis (equal); Funding acquisition (equal); Methodology (equal); Project administration (equal); Resources (equal); Software (equal); Supervision (equal); Visualization (equal); Writing-review \& editing (equal). \textbf{Inmaculada Pascual}: Funding acquisition (equal); Writing-review \& editing (equal). \textbf{Augusto Beléndez}: Funding acquisition (equal); Resources (equal). 

\section{DATA AVAILABILITY}

The data that supports the findings of this study are available from the corresponding author upon reasonable request.

\section{REFERENCES}

\nocite{*}
\bibliography{reference}

\end{document}